\documentclass[letter]{aa}
\usepackage{graphicx}
\usepackage[varg]{txfonts}
\usepackage{physics}
\usepackage{amsmath, amssymb}
\usepackage{dsfont}
\usepackage{bbm}

\usepackage{ulem}
\usepackage{xpatch}
\usepackage{hyperref}
\usepackage{import}
\usepackage{makeidx}
\usepackage{changes}
\usepackage{placeins}
\usepackage{soul}

\newcommand{\sref}[1]{Sec.~\ref{#1}}   

\newcommand{\fig}[1]{Fig.~\ref{#1}}
\newcommand{\app}[1]{Appendix~\ref{#1}}

\newcommand{\equo}[1]{Eq.~\ref{#1}}

\newcommand{\Msolpyr}{\mathrm{M_\odot~yr^{-1}}}

\newcommand{\Mdot}{\Dot{M}_\star}

\newcommand{\colout}[1]{\bgroup\markoverwith{\textcolor{#1}{\rule[.5ex]{2pt}{0.4pt}}}\ULon}

\makeatletter
\renewcommand*\aa@pageof{, page \thepage{} of \pageref*{LastPage}}

\makeatother

\makeindex

\begin{document}


%
\titlerunning{Do accretion-powered stellar winds help spin down T Tauri stars?}
\authorrunning{L.~Gehrig et al.}
\title{Do accretion-powered stellar winds help spin down T Tauri stars?} 
%
%
\author{
 Lukas~Gehrig\inst{1}, Eric Gaidos\inst{1,2}, Laura Venuti\inst{3,4}, Ann Marie Cody\inst{3}, and Neal J. Turner\inst{5}
}
\institute{
 Department of Astrophysics, University of Vienna,
 Türkenschanzstrasse 17, 1180 Vienna, Austria
 \and
 Department of Earth Sciences, University of Hawai'i at M\={a}noa, Honolulu, Hawai'i 96822 USA
 \and
 SETI Institute, 339 Bernardo Ave., Suite 200, Mountain View, CA 94043, USA
\and
Visiting Fellow, School of Physics, UNSW Science, Kensington, NSW 2052, Australia
\and
Jet Propulsion Laboratory, California Institute of Technology, 4800 Oak Grove Drive, Pasadena, California 91109, USA
}
\date{Received ....; accepted ....}

\abstract
{

How T Tauri stars remain slowly rotating while still accreting material is a long-standing puzzle.
Current models suggest that these stars may lose angular momentum through magnetospheric ejections of disk material (MEs) and accretion-powered stellar winds (APSWs).
The individual contribution of each mechanism to the stellar spin evolution, however, is unclear.
We explore how these two scenarios could be distinguished by applying stellar spin models to near-term observations.  We produce synthetic stellar populations of accreting Class~II stars with spreads in the parameters governing the spin-down processes and find that an APSW strongly affects the ratio of the disk truncation radius to the corotation radius, $\mathcal{R} = R_\mathrm{t}/R_\mathrm{co}$.
The ME and APSW scenarios are distinguished to high confidence when at least $N_\mathrm{crit}\gtrsim 250$~stars have values measured for $\mathcal{R}$.  
Newly developed lightcurve analysis methods enable measuring $\mathcal{R}$ for enough stars to distinguish the spin-down scenarios in the course of upcoming observing campaigns.
}


\maketitle


\section{Introduction}
\label{sec:intro}

The slow rotation of many T Tauri stars (with ages $\lesssim 10$~Myr and masses $\lesssim 2~\mathrm{M_\odot}$) is an unsolved problem.
The addition of angular momentum (AM) from an accreting circumstellar disk, plus contraction of the pre-main sequence star, should lead to spin-up \citep[e.g.,][]{Rebull2002}.  But surveys of star-forming regions and young clusters show the distribution of rotation periods to be little changed over the 1--10 Myr lifetime of protoplanetary disks \citep[PPD, e.g.,][]{Rebull2006, Cody2010, Venuti2017, Rebull2020, Serna2021, Smith2023}. Most stars have rotation periods between 1 and 10 days, with a modal value of $\sim$3-4 days.  It is only after PPD dissipation that stars show the expected spin-up during their evolution toward the main sequence \citep[e.g.,][]{Gallet13}.

The interaction between the star and the PPD is assumed to remove AM from the star, preventing spin-up \citep[e.g.,][]{Koenigl91}.
Current stellar spin evolution models \citep[e.g.,][]{Gallet19, Pantolmos20, Ireland21, Ireland2022} explain the loss of AM by two mechanisms. First, magnetospheric ejections \citep[MEs, ][]{Zanni13} are the result of the relative motion of the stellar magnetic field through the partially ionized accretion disk, generating a toroidal magnetic field.  The resulting additional magnetic pressure can start a cycle of inflation, reconnection, and contraction of field lines.  During this cycle, some disk material is loaded onto and ejected via magnetic field lines rooted at the disk's surface, creating a torque.  The second mechanism is accretion-powered stellar winds (APSW), where a fraction $W$ of the material accreted onto the star is ejected along open stellar magnetic field lines, removing AM \citep[e.g.,][]{Matt05APSW, Matt08, Matt2012APSW, Finnley18}. 
\cite{Zanni2011} have shown that an APSW is unlikely to be the sole mechanism that spins down a T Tauri star but the contribution of each mechanism is unclear since parameters such as $W$ and the magnetic field strength are unknown or weakly constrained.

Here we use the spin evolution model of \cite{Ireland2022} to identify differences in the multiple observables of T Tauri stars due to the contribution of an APSW.  We construct a spin population model to estimate the minimum sample size of an observed distribution to differentiate between the no-wind ($W=0$ and only MEs spin down the star) and wind scenarios ($W>0$ plus MEs).


\section{Spin population model}
\label{sec:spin_model}

\subsection{Stellar spin model}

\begin{figure}
    \centering
         \resizebox{\hsize}{!}{\includegraphics{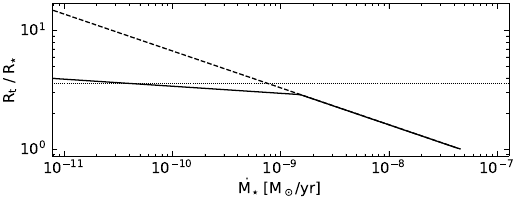}}
    \caption{
    Disk truncation radius $R_\mathrm{t}$ in units of stellar radius $R_{\star}$ vs. accretion rate $\Mdot$ for a 2 Myr-old, 0.7-$\mathrm{M_\odot}$ star with $B_\star = 0.5$~kG, and $P_\star=2$~days.  The solid line shows the expression given in \equo{eq:Rt}.
    The dashed line shows the first term of \equo{eq:Rt} and mimics (except for slightly different fitting parameters) the analytically derived expression.
    The horizontal dashed line marks the co-rotation radius.
    In the region of $\Mdot \sim 10^{-7} \Msolpyr$, the condition $R_\mathrm{t}\geq R_\star$ would be violated for the given parameters, and these values are excluded.
    }
    \label{fig:Rt}
\end{figure}

We use the model outlined in \cite{Ireland2022}, in which the external torque acting on a star consists of two components, $\Gamma_\mathrm{ext} = \Gamma_\mathrm{SDI} +  \Gamma_\mathrm{W}$, with the star-disk interaction torque, $\Gamma_\mathrm{SDI}$ and the torque due to an APSW, $\Gamma_\mathrm{W}$. 
In the picture of $\Gamma_\mathrm{SDI}$, AM can be transferred between the star and disk via the accretion of disk material and magnetic star-disk interaction (ME). Accreted material adds AM to the star, while the magnetic star-disk interaction adds or removes AM from the star, depending on the ratio of the truncation radius to the corotation radius $\mathcal{R} = R_\mathrm{t}/R_\mathrm{co}$.  
For $\mathcal{R}<1.0$~($>1.0$) the disk's inner edge rotates faster (slower) compared to the stellar surface.
The corotation radius is given by $R_\mathrm{co} = (G M_\star / \Omega_\star^2)^{1/3}$, with the gravitational constant, $G$, the stellar mass, $M_\star$, and the stellar rotation rate, $\Omega_\star$.  
The truncation radius, $R_\mathrm{t}$, is where stellar magnetic forces disrupt the disk structure.
Following \cite{Ireland2022}, we use the expression
\begin{equation}\label{eq:Rt}
    R_\mathrm{t}/R_\star = \mathrm{min}\left( K_\mathrm{t,1} \Upsilon_\mathrm{acc}^{m_\mathrm{1}},~K_\mathrm{t,2} \Upsilon^{m_\mathrm{2}} f^{m_\mathrm{3}} \right)\, ,
\end{equation}
with the disk magnetization parameter, $\Upsilon_\mathrm{acc} = B_\star^2 R_\star^2 / (4 \pi \Dot{M}_\star v_\mathrm{esc})$, the fraction of the break up speed, $f$, and the constants $K_\mathrm{t,1}= 0.772$, $K_\mathrm{t,2}= 1.36$, $m_\mathrm{1}= 0.311$, $m_\mathrm{2}= 0.0597$, and $m_\mathrm{3}= -0.261$.  The stellar radius, magnetic field, and accretion rate are denoted as $R_\star$, $B_\star$, and $\Dot{M}_\star$, and $v_\mathrm{esc}$ is the stellar escape velocity $\sqrt{2GM_\star/R_\star}$. 
While the first term in \equo{eq:Rt} is similar to the analytically derived expression for the truncation radius used in \cite{bessolaz08} or \cite{Gallet19}, the second term describes the truncation radius when it is close to the co-rotation radius, $\mathcal{R}\approx 1$.
This condition can occur if, for example, $\Mdot$ is low or $B_\star$ is large.
When $R_\mathrm{t}$ is approaching $R_\mathrm{co}$, the stellar magnetic pressure must overcome the growing disk's thermal pressure due to a pile-up of disk material just outside $R_\mathrm{co}$ in addition to the ram pressure \citep[e.g.,][]{bessolaz08, Steiner21}.
In addition, we require that $R_\mathrm{t} \geq R_\star$.
The dependence of $R_\mathrm{t}$ on $\Mdot$ is shown in \fig{fig:Rt}. 
When approaching the co-rotation radius, the thermal pressure of the disk close to $R_\mathrm{co}$ slows down the decrease of $R_\mathrm{t}$.
The transition between the two contributions of $R_\mathrm{t}$ occurs at $\mathcal{R}\approx 0.7$.
For values of $\mathcal{R}\gtrsim 0.7$, \equo{eq:Rt} predicts significantly smaller values of $R_\mathrm{t}$ compared to its analytical expression. 
We note that the exact position of the transition between the two regimes and the range where $\mathcal{R}>1$ depend on multiple parameters, for example, $B_\star$, $\Mdot$, and $P_\star$ (see \equo{eq:Rt}).
Furthermore, specific parameter combinations, for example, $\Mdot \sim 10^{-7}~\Msolpyr$, can lead to $R_\mathrm{t}<R_\star$ and are therefore excluded from this work \citep[as suggested in][]{Ireland2022}.
The exact range of accretion rates leading to $R_\mathrm{t}<R_\star$ also depends on other parameters, such as $B_\star$, $M_\star$, or the stellar age (see \equo{eq:Rt}).

\citet{Ireland2022} divide the star-disk interaction torque $\Gamma_\mathrm{SDI}$ into three states:  In State 1, for $\mathcal{R}< 0.433$, accretion and MEs spin up the star.  For values of $0.433 \leq \mathcal{R} \leq 1.0$, accretion still spins up the star, but MEs can exert a spin-down torque (State 2).  In State 3, for $\mathcal{R}> 1.0$, the spin-up torque due to accretion is negligible and the MEs spin down the star.  The expressions for the corresponding torques are:
\begin{equation}\label{eq:torque}
    \Gamma_\mathrm{SDI} = 
    \begin{cases}
    K_\mathrm{SDI,1} \Dot{M}_\star (G M_\star R_\mathrm{t})^{0.5} & \text{State~1} \\
    K_\mathrm{SDI,1} \Dot{M}_\star (G M_\star R_\mathrm{t})^{0.5}- K_\mathrm{SDI,2} B_\star^2 R_\star^6/R_\mathrm{co}^3 & \text{State~2} \\
    -K_\mathrm{SDI,2} B_\star^2 R_\star^6/R_\mathrm{t}^3 & \text{State~3} \\
    \end{cases}
\end{equation}
with the constants $K_\mathrm{SDI,1}= 0.909$ and $K_\mathrm{SDI,2}= 0.0171$.
The torque due to an accretion-powered stellar wind can be written as $\Gamma_\mathrm{W} = \Dot{M}_\mathrm{W} \Omega_\star r_\mathrm{A}^2$ \citep[e.g.,][]{Weber67} with the average Alfvén radius, $r_\mathrm{A}$, and the APSW mass loss rate $\Dot{M}_\mathrm{W} = W \Mdot$.
The Alfvén radius can be related to the open magnetic flux of the stellar wind \citep[][]{Reville2015}, yielding the expression of $\Gamma_\mathrm{W}$ used in \cite{Ireland2022}
\begin{equation}\label{eq:wind}
\begin{split}
    \Gamma_\mathrm{W} = &- K_\mathrm{A,1}^2 K_\mathrm{\phi}^{4 m_\mathrm{A}} \Dot{M}_\mathrm{W} (G M_\star R_\star)^{0.5} f^{1+4 m_\mathrm{\phi,2} m_\mathrm{A}} \left( \frac{R_\mathrm{t}}{R_\star} \right)^{4 m_\mathrm{\phi,1} m_\mathrm{A}} \\
    & \times \left( \frac{\Upsilon_\mathrm{\star}}{(1+(f/K_\mathrm{A,2})^2)^{0.5}}  \right)^{2 m_\mathrm{A}}
\end{split}
\end{equation}
with the fitting parameters $K_\mathrm{A,1} = 0.954$, $K_\mathrm{A,2} = 0.0284$, $m_\mathrm{A}=0.394$, $K_\mathrm{\phi}= 1.62$, $m_\mathrm{\phi,1}= -1.25$, $m_\mathrm{\phi,2}= 0.184$, and the magnetization parameter of the stellar wind
\begin{equation}
    \Upsilon_\mathrm{\star} = \frac{\Phi_\star^2}{4 \pi R_\star^2 \Dot{M}_\mathrm{W} v_\mathrm{esc}} \, ,
\end{equation}
with the total unsigned stellar magnetic flux, $\Phi_\star = 2 \pi R_\star^2 B_\star$.
One important aspect of this formulation is the stronger dependence of $r_\mathrm{A}$ on $R_\mathrm{t}$ and $\Mdot$ compared to other studies \citep[e.g.,][]{Gallet19}.  A more detailed comparison is shown in \app{sec:comp_wind}.
The parameter $W$ defines the mass-loss rate of the APSW and ranges up to $\sim 2\%$ \citep[][]{Browning16, Pantolmos20}.
We note that $W$ does not account for the complete ejection ratio of the inner star-disk interaction region; that can reach up to $\sim 60\%$ and includes processes such as APSW, MEs, or disk winds \citep[e.g.,][]{Watson2016, Serna2024}.

This study compares two scenarios: $W=0\%$, corresponding to no APSW, and $W=1\%$.  One key parameter of the spin model is the ratio $\mathcal{R}$ \citep[e.g.,][]{Gallet19, Pantolmos20, Ireland21, Ireland2022}.  For a given star ($M_\star$ and $R_\star$ known), $\Omega_\star = 2\pi/P_\star$ sets $R_{\rm co}$, while $R_{\rm t}$ depends on $\Dot{M}_\star$ and $B_\star$.  Furthermore, the value of $\mathcal{R}$ is a measure of the rotational state of the star; for increasing values of $\mathcal{R}$, the spin-up (spin-down) tendency decreases (increases).  Thus, $P_\star$, $\Dot{M}_\star$, and $\mathcal{R}$ define the torque components and rotational properties of a given star.

\subsection{Equilibrium state}\label{sec:equi_theo}

\begin{figure}
    \centering
         \resizebox{\hsize}{!}{\includegraphics{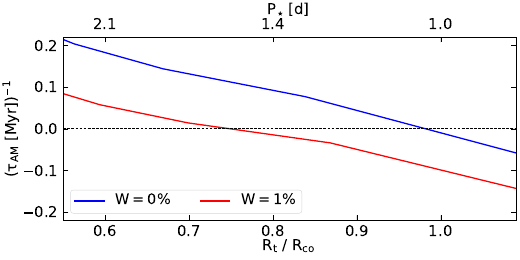}}
    \caption{
    Net torque in units of the stellar angular momentum vs $R_\mathrm{t}/R_\mathrm{co}$ and $P_\star$ for two values of the wind parameter $W$.  
    For the wind scenario ($W=1\%$), the zero-torque condition (horizontal line) is reached at lower values compared to the no-wind scenario ($W=0\%$).
    The stellar parameters are as follows: $M_\star = 0.7~\mathrm{M_\odot}$, $B_\star = 0.5$~kG, $\Mdot = 3\times10^{-9}~\Msolpyr$, and $\tau_\mathrm{age}=2$~Myr.
    }
    \label{fig:schem_plot}
\end{figure}

As described in \cite{Ireland2022} and \cite{Mueller2024}, the dependence of $\Gamma_{\rm SDI}$ on rotation is stabilizing and can cause an accretion disk-hosting star to evolve toward an equilibrium, or zero-torque, rotational state within a timescale $ \tau_\mathrm{AM} = J_\star / \Gamma_\mathrm{ext} \lesssim 1-2$~Myr, where $J_\star$ is the stellar angular momentum.  An extra, positive external torque causes the star to spin up, the corotation radius to contract, $\mathcal{R}$ to increase, and $\Gamma_{\rm SDI}$ to decrease.  Conversely, a negative external torque will cause $\Gamma_{\rm SDI}$ to increase. 
This behavior is illustrated in \fig{fig:schem_plot}.   Increasing $\mathcal{R}$, which corresponds to faster rotation, causes the net torque to decrease, causing the star to spin down, $R_{\rm co}$ to increase, and stabilizing $\mathcal{R}$ and hence $P_\star$.
In the absence of an APSW (blue line), the equilibrium state is based on the relation between the spin-up torque due to accretion and the spin-down torque of MEs.  Thus, the equilibrium is reached for values of $0.433 \leq \mathcal{R} \leq 1.0$ (State 2, see \equo{eq:torque}).
Because an APSW removes AM from the star, the equilibrium state is reached at lower values of $\mathcal{R}$ and higher values of $P_\star$.
To evolve toward a spin equilibrium, our model assumes a constant magnetic field strength and does not incorporate a relation between $B_\star$ and $P_\star$.
Possible short-term variations of $B_\star$ are discussed in \sref{sec:uncert}.
The effect of the changing moment of inertia of the contracting star can be neglected since the contraction is slow compared to $\tau_\mathrm{AM}$ for the stellar masses and ages studied in this work \citep[e.g.,][]{Ireland2022}.

\begin{figure}
    \centering
         \resizebox{\hsize}{!}{\includegraphics{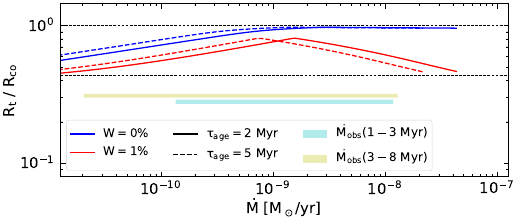}}
    \caption{
    Equilibrium state values of $R_\mathrm{t}/R_\mathrm{co}$ over stellar accretion rates.
    We compare the no-wind (blue lines) with the wind (red lines) scenario for two stellar ages, $\tau_\mathrm{age}=2~\mathrm{Myr}$ and $\tau_\mathrm{age}=5~\mathrm{Myr}$.
    For comparison, the observed range of stellar accretion rates is shown for stellar ages between 1~Myr and 8~Myr (cyan and yellow lines).
    The stellar mass is $0.7~\mathrm{M_\odot}$ and the magnetic field strength is 0.5~kG.
    The horizontal dashed lines enclose State 2 (\equo{eq:torque}).
    In the region of $\Mdot \sim 10^{-7}  \Msolpyr$, the condition $R_\mathrm{t}\geq R_\star$ would be violated, and these values are excluded.
    }
    \label{fig:theo_age}
\end{figure}

The equilibrium state for the no-wind scenario with $W=0\%$ and the wind scenario $W=1\%$ is shown in \fig{fig:theo_age} over a range of stellar accretion rates typical for T Tauri stars.
We consider stellar ages $\tau_\mathrm{age}$ of 2~Myr (solid lines) and 5~Myr (dashed lines).  
For comparison, we show the observed distribution of stellar accretion rates, $\Dot{M}_\mathrm{obs}$, at 1--3 and 3--8 Myr from \citet{Betti2023}.
In the range of $\Dot{M}_\mathrm{obs}$, the APSW reduces the equilibrium values of $\mathcal{R}$ by 15-35\%.
The dependence of $r_\mathrm{A}$ on $R_\mathrm{t}$ and $\Mdot$ results in an increasing efficiency of the APSW with higher accretion rates.
Analogous to the no-wind scenario, the equilibrium values remain between $0.433 \leq \mathcal{R} \leq 1.0$ (State 2), indicating that the APSW alone cannot counteract the spin-up torques \citep[e.g.,][]{Zanni2011}.
The break in the equilibrium values of $\mathcal{R}$ for $W=1\%$ can be explained by the transition of $R_\mathrm{t}$ close to $R_\mathrm{co}$ (see \equo{eq:Rt} and \fig{fig:Rt}) and also depends on $B_\star$.
In contrast, the stellar mass dependence of the equilibrium state is weak compared to the effect of an APSW (see also \fig{fig:delta}).

The equilibrium state shown in  Fig.~\ref{fig:theo_age} is based on fixed stellar and disk parameters.  However, the parameters do not stay constant but evolve over Myr.  For example, the stellar radius contracts during pre-main sequence evolution, and the accretion rate decreases during the Class II phase.  As a result, the equilibrium value of $\mathcal{R}$ slowly changes over time.
The star, however, remains close to its equilibrium state during its evolution once it is reached \citep[e.g.,][]{Ireland2022,Serna2024}.

\subsection{Effect of short-term variability}\label{sec:uncert}

Some stellar properties also exhibit short-term variability (timescales of $\tau_\mathrm{var} \lesssim$~years), especially $\Mdot$ and $B_\star$.  As a result, a star in its equilibrium state is subjected to a variation in its parameters and can experience a spin-up or spin-down torque, introducing scatter in the stellar properties.  
Typical literature values for these variations are 0.40~dex for $\Dot{M}_\star$ \citep[e.g.,][]{Manara22} and 0.48~dex for $B_\star$ \citep[e.g.,][]{Johnstone14, Reiners22}.  

\begin{figure}
    \centering
         \resizebox{\hsize}{!}{\includegraphics{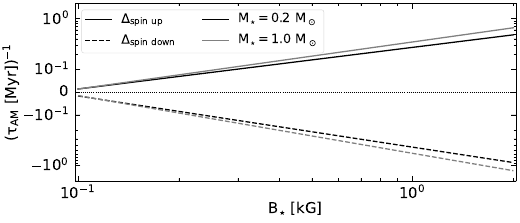}}
    \caption{
    Effect of a short-term deviation from the equilibrium state (as described in \sref{sec:uncert}).
    The resulting torques that act on the star, in units of the stellar angular momentum, $\tau_\mathrm{AM}^{-1}$, are shown over a range of equilibrium magnetic field strengths.
    The spin-up (down) torques are shown in solid (dashed) lines.  
    The stellar age is 2~Myr.
    }
    \label{fig:delta}
\end{figure}

We use this variation to calculate the torque that acts on a star when $\Mdot$ and $B_\star$ deviate from their equilibrium values.  
In \fig{fig:delta}, the respective torques are shown over a range of (equilibrium) stellar magnetic field strengths for 2~Myr old stars and $W=0\%$.  
Starting from the equilibrium values, we randomly vary $\Mdot$ and $B_\star$ within the variation range and calculate the torques on the star assuming that it still rotates at the equilibrium rate ($\tau_\mathrm{var} \ll \tau_\mathrm{AM}$).  
This process is repeated 1000 times, and the solid and dashed lines mark the $1\sigma$ range.  Spin-up (spin-down) torques, $\Delta_\mathrm{spin~up}$ ($\Delta_\mathrm{spin~down}$), are marked as solid (dashed) lines. 
The magnitude of the scatter due to variability, $\Delta_\mathrm{spin~up}$ and $\Delta_\mathrm{spin~down}$, is not significantly changed by the appearance of an APSW.
When increasing the stellar age to 5~Myr, the values of $\Delta_\mathrm{spin~up}$ and $\Delta_\mathrm{spin~down}$ decrease by a factor of $\approx 2/3$ due to the smaller stellar radii for older stars and the resulting smaller torques \citep[see \equo{eq:torque} and, e.g.,][]{Serna2024}.  Since these variations appear stochastic, in our simulations we assume a star may deviate randomly from its equilibrium state within $2\Delta_\mathrm{spin~up}$ and $2\Delta_\mathrm{spin~down}$ when calculating the spin population model.

\subsection{Schematic of the stellar population model}\label{sec:pop_mod}

The main assumptions of our stellar population model (SPM) are: 1) Most stars with ages $\tau_\mathrm{age} \gtrsim 1-2$~Myr have reached their equilibrium state. 2) In their equilibrium state, there is a significant difference in the value of $\mathcal{R}$ when comparing the no-wind and wind scenarios (see \fig{fig:theo_age}).  3) Due to variation in their accretion and magnetic fields, stars can deviate from their equilibrium state within $2\Delta_\mathrm{spin~up}$ and $2\Delta_\mathrm{spin~down}$.

We generate a population of stars in three steps. 
First, we select a stellar mass between 0.2 and 1.0~$\mathrm{M_\odot}$ according to the initial mass function of \cite{Chabrier2014}.  The stellar radius is chosen based on the isochrones of \cite{Baraffe15} for a selected age.
In the second step, the accretion rate is chosen based on the relations presented in \cite{Betti2023}, allowing a scatter of two standard deviations.  Specifically, for $\tau_\mathrm{age}=2$~Myr, we use $\log\Dot{M}_\star = 2.12 \times \log M_\star -8.11$ and for $\tau_\mathrm{age}=5$~Myr, $\log\Dot{M}_\star = 2.43 \times \log M_\star -8.21$.  The $1~\sigma$ dispersions are 0.76 and 1.06~dex, respectively.
In the third step, we randomly choose a magnetic field strength from a uniform distribution between 0.1 and 2.0~kG.
For stellar rotation periods between 0.3 and 50~days \citep[the observed range, e.g.,][]{Smith2023} we calculate $R_\mathrm{t}$, $R_\mathrm{co}$, and $\Gamma_\mathrm{ext}$ according to \cite{Ireland2022} and evaluate if $2\Delta_\mathrm{spin~down}\leq \tau_\mathrm{AM}^{-1} \leq 2\Delta_\mathrm{spin~up}$.
We note that this assumes that the variability in $B_\star$ and $\Mdot$ is uncorrelated.
From all periods that fulfill this torque balance condition, we randomly choose the star's rotation period.
We discuss the sensitivity of our results to the choices of these distributions in \sref{sec:sens}.

Before showing the results of the SPM, we highlight the importance of the parameter $\mathcal{R}$ for distinguishing between the no-wind and wind scenarios. The equilibrium state and the torques due to variations in the stellar parameters strongly depend on the magnetic field strength.
For a given accretion rate and rotation period, the value $\mathcal{R}$ is directly related to $B_\star$, independent of $W$ (see \equo{eq:Rt}).
Without a measurement of $\mathcal{R}$, for example, with only measurements of $\Mdot$ and $P_\star$, the values of $\mathcal{R}$ and $B_\star$ are degenerate (covariant), preventing quantitative determinations.


\section{Results}
\label{sec:results}

We assess the potential for differentiating between the no-wind and wind scenarios using the distributions of the SPM's key parameters $P_\star$, $\Mdot$, and $\mathcal{R}$.
As described in \sref{sec:pop_mod}, we generate and compare mock distributions of the SPM parameters to estimate the sample size required to achieve a $3\sigma$ level of statistical confidence.
The mock distributions include expected observational uncertainties, increasing the scatter.
The measurement of $\Dot{M}_\star$ is subject to uncertainties of 0.35~dex \citep[e.g.,][]{Alcala2017, Manara22}.  
Stellar rotation periods can be observed with uncertainties of only a few percent \citep[e.g.,][]{Smith2023}.  
Thus, the uncertainty in $\mathcal{R}$ depends largely on the uncertainty in $R_\mathrm{t}$, which we take to be $0.2R_\star$ (Venuti et al., in prep.; see Sec.~\ref{sec:obs}).

\begin{figure}
    \centering
         \resizebox{\hsize}{!}{\includegraphics{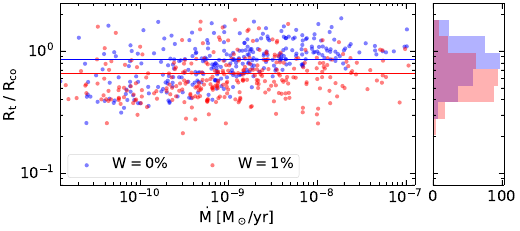}}
    \caption{
    Distribution of the spin population model (SPM) on the $\mathcal{R}-\Mdot$ plane for the wind and no-wind scenarios.  The solid lines are the mean values.  The histograms of the parameter $\mathcal{R}$ are shown on the right.  The stellar age is $\tau_\mathrm{age}=2$~Myr, and each distribution contains 300 stars.
    }
    \label{fig:slope_hist}
\end{figure}

\begin{figure}
    \centering
         \resizebox{\hsize}{!}{\includegraphics{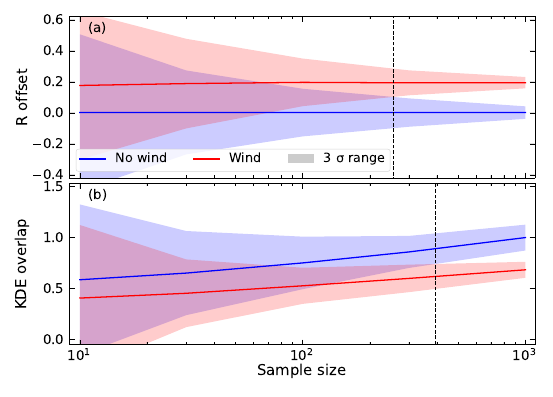}}
    \caption{
    Statistical differentiation between the no-wind and wind scenarios.
    The offset values of $\mathcal{R}$ and the normed value of the overlap of the integrated KDE products are shown in Panels (a) and (b).
    The solid lines are the mean values after 1000 realizations of the SPM. 
    The shaded areas indicate the $3\sigma$ range.
    The stellar age is 2~Myr.
    The vertical dashed lines show the values of $N_\mathrm{crit}$.
    }
    \label{fig:KDE}
\end{figure}

One way to distinguish between the no-wind and wind scenarios is the offset between the mean values of $\mathcal{R}$ for different wind scenarios (see \fig{fig:slope_hist}).
There is a distinct offset (in the mean values) between populations of stars with APSWs and without them (\fig{fig:slope_hist}).
The histogram on the right of \fig{fig:slope_hist} shows the distribution of the parameter $\mathcal{R}$.  
For the no-wind and wind scenarios, the maxima of the distributions are located at $\approx 0.8-0.9$ and $\approx 0.6-0.7$. 
Although all equilibrium values of $\mathcal{R}$ are located within $0.433 \leq \mathcal{R} \leq 1.0$ (State 2), the effects of short-term variability (\sref{sec:uncert}) are responsible for stars shifted out of their equilibrium into State 1 or State 3.
These stars are spinning up (if located in State 1) or spinning down (if located in State 3) toward their equilibrium in State 2.
To compare the scenarios, we generate a sample of 1000 stars assuming $W=0\%$ as reference.
Then, for different sample sizes (ranging between 10 and 1000 stars), we generate populations for $W=0\%$ and $W=1\%$ and calculate the offsets between the mean values of $\mathcal{R}$ of the respective distributions and the reference sample.
This process is repeated 1000 times.
The resulting mean values and $3\sigma$ ranges of the offsets vs. sample size are shown in Panel (a) of \fig{fig:KDE} for $\tau_\mathrm{age}=2$~Myr. 
To distinguish between the wind and no-wind scenarios on a high confidence ($3\sigma$) level, we require a sample size of $N_\mathrm{crit} \ge 255$ stars. 
For older stars ($\tau_\mathrm{age}=5$~Myr), $N_\mathrm{crit}$ reduces to 218. 
The smaller scatter due to variability among older stars (see \sref{sec:uncert}) is responsible for this lower value of $N_\mathrm{crit}$ for older stars.

Alternatively, one can distinguish between the scenarios by comparing the overall parameter distributions, e.g., using a kernel density estimator (KDE), an estimate of the probability density function.  
We do not use the Kolmogorov-Smirnov test because its sensitivity decreases towards the tails of the distributions \citep[e.g.,][]{Lanzante2021}.
For comparing two distributions of the SPM parameters $P_\star$, $\Dot{M}_\star$, and $\mathcal{R}$, the respective KDEs are multiplied and the result is integrated over the whole parameter space.  The larger the value of this integral, the more the KDEs overlap, and the more similar the distributions are.   As a reference and for normalization, we use a population with 1000~stars and assume $W=0\%$.  Similar to the previous method, this comparison is repeated 1000 times for different sample sizes, assuming $W=0\%$ and $W=1\%$. 
The distribution of the metric is shown in Panel (b) of \fig{fig:KDE}.
For this method, we find $N_\mathrm{crit}=395$.
Similarly to the comparison of the offsets, $N_\mathrm{crit}$ decreases to 323 for $\tau_\mathrm{age}=5$~Myr.

\section{Discussion and Conclusion}
\label{sec:discussion}


\subsection{Sensitivity tests}\label{sec:sens}

In the SPM, we randomly choose the magnetic field strength and the stellar rotation period from the expected variation since the actual distributions of these parameters are poorly constrained.  
Now, we explore how the results depend on the way these parameters are chosen.
First, we change the distribution of $B_\star$.
While the range is still between 0.1 and 2.0~kG, the distribution is centered at 1.5~kG with a deviation of 0.3~kG, shifting the magnetic field strengths to higher values.
As a result, the theoretical difference between the wind and no-wind scenarios decreases due to more effective MEs, and the expected variation increases (see \fig{fig:delta}).  
This is reflected by an increase in $N_\mathrm{crit}$ by 30\%.  

Next, we modify the torque variation due to short-term variability (see \sref{sec:uncert}).   The torque can still vary between $2\Delta_\mathrm{spin~up}$ and $2\Delta_\mathrm{spin~down}$ but it is chosen from a normal distribution located at the center of $\Delta_\mathrm{spin~up}$ and $\Delta_\mathrm{spin~down}$ with a deviation of the distance between the center $\Delta_\mathrm{spin~up}$.  As a result, the rotation periods are located closer to the equilibrium state, and the value of $N_\mathrm{crit}$ decreases by 18\%.

In this work, we have restricted the wind parameter $W$ to two values, 0\% and 1\%.
It is, however, likely that the values of $W$ are distributed along a certain range.
We want to discuss how our results would be affected if $W$ is distributed uniformly between 0\% and 2\%.
Although the variation is mostly unaffected by the value of $W$ (see \sref{sec:uncert}), the equilibrium value of $\mathcal{R}$ would change (see \sref{sec:equi_theo}).
For lower values of $W$, the equilibrium states of the no-wind and wind scenarios are located closer together, resulting in more similar distributions and an increase in $N_\mathrm{crit}$.
Similarly, for higher values of $W$, $N_\mathrm{crit}$ decreases.
As a result, using the aforementioned uniform distribution of $W$ between 0\% and 2\%, the mean values of the distributions should remain approximately constant while the scatter increases.
Thus, $N_\mathrm{crit}$ would also increase but should remain the same order of magnitude.
We note that this estimation can change based on the assumed distribution of $W$.

Finally, we want to discuss the dependence of our results on the APSW model. 
In this work, we utilize the model presented in \cite{Ireland2022}. 
Compared to other APSW models, for example, the model presented in \cite{Matt2012APSW}, which is also used in \cite{Gallet19}, the APSW model of \cite{Ireland2022} includes the effects of an accretion disk leading to stronger dependence of $r_\mathrm{A}$ on $\Mdot$.
In \app{sec:comp_wind}, we compare the two wind models in more detail.
The average difference between the APSW torque models is relatively small (\fig{fig:WG19}) for $10^{-10}~\mathrm{M_\odot/yr} \leq \Mdot \leq 10^{-8}~\mathrm{M_\odot/yr}$, which is a representative range for the majority of young stars \citep[e.g.,][]{Betti2023}.
The value of $N_\mathrm{crit}$ increases by $\approx 20\%$ when using the wind model from \cite{Gallet19}.
We note that the parameter ranges used to formulate the APSW models do not completely cover the parameters used in this study (see \app{sec:paraIreland}).
For specific parameter combinations close to the tails of observed distributions (e.g., very high accretion rates, $\Mdot\sim 10^{-7}~\Msolpyr$, and low magnetic field strengths, $B_\star\sim0.1$~kG), the relations of the stellar spin model are limited.
Since the model presented in \cite{Ireland2022} includes the effects of the accretion disk on the Alfvèn radius and the average difference compared to other models are small for the majority of observed stars, we have chosen their description to model the APSW torque, knowing that for certain parameter ranges, the relation might be updated in the course of future studies.

\subsection{Availability of observations}\label{sec:obs}

Several studies have collected an impressive number of parameters for YSOs, including mass, radius, rotation period, and accretion rate \citep[e.g.,][]{Betti2023, Smith2023}.
On the other hand, the large-scale magnetic field strength and the location of the truncation radius are poorly constrained by observations.
Currently, there are a few dozen observations of magnetic field strengths of accreting stars \citep[e.g.,][]{Johnstone14, Reiners22,Finociety2023,Flores2024,Donati2024a,Donati2024b} and about as many $R_\mathrm{t}$ estimates derived from constraining the size of the Br$\gamma$ emitting region around young stars with infrared interferometry \citep{Eisner2014,Gravity2023}.
The ensemble of these observations is not sufficient to test the SPM adequately.

However, the 1-dimensional model of time-dependent accretion in young stars developed by \citet{Robinson2021} has enabled the development of the first prescription that directly links light curve morphology and color variability of accreting young stars to the structure of magnetospheric accretion. The theoretical underpinning of this prescription is the following: the kinetic energy of the accretion flow deposited onto the star depends on the distance it traveled from its starting point $R_\mathrm{t}$. 
This dependence is reflected in the emission spectrum of the resulting accretion shock, which can be probed observationally by measuring the color of the ultraviolet (UV) excess. 
A pilot observational program to validate this prescription is presented in Venuti et al. (2025, in prep.), who employ simultaneous monitoring data of accreting young stars at near-UV and optical wavelengths to populate the stellar loci on color-magnitude diagrams and fit them with the \citet{Robinson2021} grid of predicted color traces to identify the best-fitting magnetospheric parameters.
Derived independently of other observational parameters, the value of $R_\mathrm{t}$ in combination with $\Dot{M}_\star$ and $P_\star$ can also provide estimates of $B_\star$ (see \equo{eq:Rt}).
Thus, the large-scale application of the technique described in Venuti et al. (2025, in prep.) could provide a significant number of $R_\mathrm{t}$ and $B_\star$ measurements, thereby allowing valuable insights into stellar spin-down mechanisms.

\subsection{Accuracy of the stellar spin model}\label{sec:limits}

The main limitation of the SPM lies in the spin model itself.
In their study, \cite{Ireland2022} utilize 2.5-dimensional simulations of the star-disk interaction region.
Based on these simulations, with runs over several dynamical timescales ($\lesssim 50~\mathrm{P_\star}$), relations for the torque acting on the star are extracted \citep[see also][]{Pantolmos20} and extrapolated over a large parameter space and the lifetime of PPDs \citep[e.g.,][]{Ireland21}.
While producing correct qualitative correlations, the absolute results of the spin model are subject to uncertainties.

Our study demonstrates the sensitivity of the observable $\mathcal{R}$ to the presence of an APSW such that with sufficient observations, different stellar spin-down mechanisms, for example, wind models, can be distinguished.  However, they also highlight certain limitations of our model.  One is the predicted stellar rotation period distribution.  While the range of observed periods between $\sim 1$ and $\sim 10$~days is reproduced, the peak in the distribution is located at 1--3~days, compared to the observed maximum at 3--4~days \citep[e.g.,][]{Smith2023}.
The value of the parameter $K_\mathrm{t,1}$ used in the formulation of the truncation radius \cite{Ireland2022} is smaller by a factor of 0.67 compared to the values in \cite{Gallet19} and \cite{Pantolmos20}.  As a result, for a given value of $\mathcal{R}$, the co-rotation radius is smaller by the same amount, and this means faster rotation, with periods shorter by a factor of 0.55.  This could at least partly explain the discrepancy in the rotation period distribution of our SPM compared to observations.  
Another aspect that varies between spin models is the amount of angular momentum added to the star due to accretion.
Magnetic torques and outflows, such as MEs, can remove AM of the disk and at $R_\mathrm{t}$, the disk rotates at velocities smaller than Keplerian \citep[e.g.,][]{Gallet19}.
The closer the disk rotates to the Keplerian value, the more AM it carries, and the greater the spin-up of the star.  Disks rotate at 0.4 and 0.9 times the Keplerian value in the \citet{Gallet19} and \citet{Ireland2022} models, respectively, thus more AM is transferred onto stars in our SPM.

Our model assumes that the stellar magnetic field is, besides short-term variations (\sref{sec:uncert}), roughly independent of other parameters such as the rotation period.  
We base this assumption on the small Rossby numbers of T~Tauri stars \citep[$Ro\lesssim0.1$, e.g.,][]{Briggs07, Johns-Krull2007}.
They usually are located in the ``saturation" regime of dynamo theory where the dependence of the magnetic field strength on the rotation period is expected to be weak \citep[$B_\star \propto P_\star^{-0.16}$, e.g.,][]{Reiners22} and small compared short-term variation (cycling) observed in $B_\star$ (\sref{sec:uncert}).

\subsection{Concluding remarks}

The relative importance of accretion-powered stellar winds compared to, for example, magnetospheric ejections in the spin-down of stars during the T~Tauri phase is unclear.
The primary purpose of this letter is to describe a method that can evaluate the contribution of APSWs using observations of an appropriate population of Class II young stellar objects.  
One difference between the spin-down mechanisms manifests itself in the distributions of $\mathcal{R} = R_\mathrm{t}/R_\mathrm{co}$.  Assuming stars during this phase are close to equilibrium (zero-torque state), our stellar population model predicts a systematic shift in the equilibrium state as an APSW removes angular momentum and reduces $\mathcal{R}$ compared to the no-wind case.  Montecarlo simulations using our SPM predict that approximately $N_\mathrm{crit}\gtrsim 250$~stars (depending on the age) with simultaneous measurements of truncation radius $R_\mathrm{t}$, rotation period $P_\star$, and accretion rate $\Mdot$ are sufficient to differentiate between the no-wind and wind scenarios at a $3\sigma$ confidence level.  
The sensitivity of the value of $N_\mathrm{crit}$ on the range of $B_\star$, short-term variability, and the description of the APSW torque is smaller than $\sim30\%$.
Our work demonstrates that the value of $N_\mathrm{crit}$ depends on the used spin model and statistical evaluation, but ranges in the order of several hundred stars.
While current observations are inadequate for such a test, future, uniform surveys of larger samples and application of a spectro-photometric technique to infer $R_t$ from stellar emission alone could prove fruitful.


\begin{acknowledgements}
We thank the anonymous referee for constructive feedback that helped to improve the manuscript.
EG was supported by NASA Award  80NSSC19K0587 (Astrophysics Data Analysis Program) and NSF Award 2106927
(Astronomy \& Astrophysics Research Program) as a Gauss Professor at the University of G\"{o}ttingen by the Nieders\"{a}chsische Akaemie der Wissenschaften.
NJT acknowledges that this work was carried out in part at the Jet Propulsion Laboratory, California Institute of Technology, under contract 80NM00018D0004 with NASA.
 
\end{acknowledgements}


\bibliographystyle{aa}
\bibliography{main}

\newcommand{\noop}[1]{}
\begin{thebibliography}{44}
\expandafter\ifx\csname natexlab\endcsname\relax\def\natexlab#1{#1}\fi

\bibitem[{{Alcal{\'a}} {et~al.}(2017){Alcal{\'a}}, {Manara}, {Natta}, {Frasca}, {Testi}, {Nisini}, {Stelzer}, {Williams}, {Antoniucci}, {Biazzo}, {Covino}, {Esposito}, {Getman}, \& {Rigliaco}}]{Alcala2017}
{Alcal{\'a}}, J.~M., {Manara}, C.~F., {Natta}, A., {et~al.} 2017, \aap, 600, A20

\bibitem[{{Baraffe} {et~al.}(2015){Baraffe}, {Homeier}, {Allard}, \& {Chabrier}}]{Baraffe15}
{Baraffe}, I., {Homeier}, D., {Allard}, F., \& {Chabrier}, G. 2015, \aap, 577, A42

\bibitem[{{Bessolaz} {et~al.}(2008){Bessolaz}, {Zanni}, {Ferreira}, {Keppens}, \& {Bouvier}}]{bessolaz08}
{Bessolaz}, N., {Zanni}, C., {Ferreira}, J., {Keppens}, R., \& {Bouvier}, J. 2008, \aap, 478, 155

\bibitem[{{Betti} {et~al.}(2023){Betti}, {Follette}, {Ward-Duong}, {Peck}, {Aoyama}, {Bary}, {Dacus}, {Edwards}, {Marleau}, {Mohamed}, {Palmo}, {Plunkett}, {Robinson}, \& {Wang}}]{Betti2023}
{Betti}, S.~K., {Follette}, K.~B., {Ward-Duong}, K., {et~al.} 2023, \aj, 166, 262

\bibitem[{{Briggs} {et~al.}(2007){Briggs}, {G{\"u}del}, {Telleschi}, {Preibisch}, {Stelzer}, {Bouvier}, {Rebull}, {Audard}, {Scelsi}, {Micela}, {Grosso}, \& {Palla}}]{Briggs07}
{Briggs}, K.~R., {G{\"u}del}, M., {Telleschi}, A., {et~al.} 2007, \aap, 468, 413

\bibitem[{{Browning} {et~al.}(2016){Browning}, {Weber}, {Chabrier}, \& {Massey}}]{Browning16}
{Browning}, M.~K., {Weber}, M.~A., {Chabrier}, G., \& {Massey}, A.~P. 2016, \apj, 818, 189

\bibitem[{{Chabrier} {et~al.}(2014){Chabrier}, {Hennebelle}, \& {Charlot}}]{Chabrier2014}
{Chabrier}, G., {Hennebelle}, P., \& {Charlot}, S. 2014, \apj, 796, 75

\bibitem[{Cody \& Hillenbrand(2010)}]{Cody2010}
Cody, A.~M. \& Hillenbrand, L.~A. 2010, The Astrophysical Journal Supplement Series, 191, 389

\bibitem[{{Donati} {et~al.}(2024{\natexlab{a}}){Donati}, {Cristofari}, {Alencar}, {K{\'o}sp{\'a}l}, {Bouvier}, {Moutou}, {Carmona}, {Gregorio-Hetem}, {M{\'e}nard}, {Artigau}, {Doyon}, {Takami}, {Shang}, {Dias do Nascimento}, {M{\'e}nard}, {Gaidos}, \& {SPIRou Science Team}}]{Donati2024b}
{Donati}, J.~F., {Cristofari}, P.~I., {Alencar}, S.~H.~P., {et~al.} 2024{\natexlab{a}}, \mnras, 535, 3363

\bibitem[{{Donati} {et~al.}(2024{\natexlab{b}}){Donati}, {Finociety}, {Cristofari}, {Alencar}, {Moutou}, {Delfosse}, {Fouqu{\'e}}, {Arnold}, {Baruteau}, {K{\'o}sp{\'a}l}, {M{\'e}nard}, {Carmona}, {Grankin}, {Takami}, {Artigau}, {Doyon}, {H{\'e}brard}, \& {the SPIRou science team}}]{Donati2024a}
{Donati}, J.~F., {Finociety}, B., {Cristofari}, P.~I., {et~al.} 2024{\natexlab{b}}, \mnras, 530, 264

\bibitem[{{Eisner} {et~al.}(2014){Eisner}, {Hillenbrand}, \& {Stone}}]{Eisner2014}
{Eisner}, J.~A., {Hillenbrand}, L.~A., \& {Stone}, J.~M. 2014, \mnras, 443, 1916

\bibitem[{{Finley} \& {Matt}(2018)}]{Finnley18}
{Finley}, A.~J. \& {Matt}, S.~P. 2018, \apj, 854, 78

\bibitem[{{Finociety} {et~al.}(2023){Finociety}, {Donati}, {Grankin}, {Bouvier}, {Alencar}, {M{\'e}nard}, {Ray}, {K{\'o}sp{\'a}l}, \& {consortium}}]{Finociety2023}
{Finociety}, B., {Donati}, J.~F., {Grankin}, K., {et~al.} 2023, \mnras, 520, 3049

\bibitem[{{Flores} {et~al.}(2024){Flores}, {Connelley}, {Reipurth}, {Boogert}, \& {Doppmann}}]{Flores2024}
{Flores}, C., {Connelley}, M.~S., {Reipurth}, B., {Boogert}, A., \& {Doppmann}, G. 2024, \apj, 972, 149

\bibitem[{{Gallet} {et~al.}(2019){Gallet}, {Zanni, C.}, \& {Amard, L.}}]{Gallet19}
{Gallet}, {Zanni, C.}, \& {Amard, L.} 2019, A\&A, 632, A6

\bibitem[{{Gallet} \& {Bouvier}(2013)}]{Gallet13}
{Gallet}, F. \& {Bouvier}, J. 2013, \aap, 556, A36

\bibitem[{{Gravity Collaboration} {et~al.}(2023){Gravity Collaboration}, {Wojtczak}, {Labadie}, {Perraut}, {Tessore}, {Soulain}, {Ganci}, {Bouvier}, {Dougados}, {Al{\'e}cian}, {Nowacki}, {Cozzo}, {Brandner}, {Caratti O Garatti}, {Garcia}, {Garcia Lopez}, {Sanchez-Bermudez}, {Amorim}, {Benisty}, {Berger}, {Bourdarot}, {Caselli}, {Cl{\'e}net}, {de Zeeuw}, {Davies}, {Drescher}, {Duvert}, {Eckart}, {Eisenhauer}, {Eupen}, {F{\"o}rster-Schreiber}, {Gendron}, {Gillessen}, {Grant}, {Grellmann}, {Hei{\ss}el}, {Henning}, {Hippler}, {Horrobin}, {Hubert}, {Jocou}, {Kervella}, {Lacour}, {Lapeyr{\`e}re}, {Le Bouquin}, {L{\'e}na}, {Lutz}, {Mang}, {Ott}, {Paumard}, {Perrin}, {Scheithauer}, {Shangguan}, {Shimizu}, {Spezzano}, {Straub}, {Straubmeier}, {Sturm}, {van Dishoeck}, {Vincent}, \& {Widmann}}]{Gravity2023}
{Gravity Collaboration}, {Wojtczak}, J.~A., {Labadie}, L., {et~al.} 2023, \aap, 669, A59

\bibitem[{{Ireland} {et~al.}(2022){Ireland}, {Matt}, \& {Zanni}}]{Ireland2022}
{Ireland}, L.~G., {Matt}, S.~P., \& {Zanni}, C. 2022, \apj, 929, 65

\bibitem[{{Ireland} {et~al.}(2021){Ireland}, {Zanni}, {Matt}, \& {Pantolmos}}]{Ireland21}
{Ireland}, L.~G., {Zanni}, C., {Matt}, S.~P., \& {Pantolmos}, G. 2021, \apj, 906, 4

\bibitem[{Johns-Krull(2007)}]{Johns-Krull2007}
Johns-Krull, C.~M. 2007, The Astrophysical Journal, 664, 975

\bibitem[{{Johnstone} {et~al.}(2014){Johnstone}, {Jardine}, {Gregory}, {Donati}, \& {Hussain}}]{Johnstone14}
{Johnstone}, C.~P., {Jardine}, M., {Gregory}, S.~G., {Donati}, J.~F., \& {Hussain}, G. 2014, \mnras, 437, 3202

\bibitem[{{Koenigl}(1991)}]{Koenigl91}
{Koenigl}, A. 1991, \apjl, 370, L39

\bibitem[{Lanzante(2021)}]{Lanzante2021}
Lanzante, J.~R. 2021, International Journal of Climatology, 41, 6314

\bibitem[{{Manara} {et~al.}(2022){Manara}, {Ansdell}, {Rosotti}, {Hughes}, {Armitage}, {Lodato}, \& {Williams}}]{Manara22}
{Manara}, C.~F., {Ansdell}, M., {Rosotti}, G.~P., {et~al.} 2022, arXiv e-prints, arXiv:2203.09930

\bibitem[{{Matt} \& {Pudritz}(2005)}]{Matt05APSW}
{Matt}, S. \& {Pudritz}, R.~E. 2005, \apjl, 632, L135

\bibitem[{{Matt} \& {Pudritz}(2008)}]{Matt08}
{Matt}, S. \& {Pudritz}, R.~E. 2008, \apj, 678, 1109

\bibitem[{Matt {et~al.}(2012)Matt, MacGregor, Pinsonneault, \& Greene}]{Matt2012APSW}
Matt, S.~P., MacGregor, K.~B., Pinsonneault, M.~H., \& Greene, T.~P. 2012, The Astrophysical Journal Letters, 754, L26

\bibitem[{{Mueller} {et~al.}(2024){Mueller}, {Johns-Krull}, {Stassun}, \& {Dixon}}]{Mueller2024}
{Mueller}, M.~A., {Johns-Krull}, C.~M., {Stassun}, K.~G., \& {Dixon}, D.~M. 2024, \apj, 964, 1

\bibitem[{{Pantolmos} {et~al.}(2020){Pantolmos}, {Zanni}, \& {Bouvier}}]{Pantolmos20}
{Pantolmos}, G., {Zanni}, C., \& {Bouvier}, J. 2020, \aap, 643, A129

\bibitem[{{Rebull} {et~al.}(2020){Rebull}, {Stauffer}, {Cody}, {Hillenbrand}, {Bouvier}, {Roggero}, \& {David}}]{Rebull2020}
{Rebull}, L.~M., {Stauffer}, J.~R., {Cody}, A.~M., {et~al.} 2020, \aj, 159, 273

\bibitem[{Rebull {et~al.}(2006)Rebull, Stauffer, Megeath, Hora, \& Hartmann}]{Rebull2006}
Rebull, L.~M., Stauffer, J.~R., Megeath, S.~T., Hora, J.~L., \& Hartmann, L. 2006, The Astrophysical Journal, 646, 297

\bibitem[{{Rebull} {et~al.}(2002){Rebull}, {Wolff}, {Strom}, \& {Makidon}}]{Rebull2002}
{Rebull}, L.~M., {Wolff}, S.~C., {Strom}, S.~E., \& {Makidon}, R.~B. 2002, \aj, 124, 546

\bibitem[{{Reiners} {et~al.}(2022){Reiners}, {Shulyak}, {K{\"a}pyl{\"a}}, {Ribas}, {Nagel}, {Zechmeister}, {Caballero}, {Shan}, {Fuhrmeister}, {Quirrenbach}, {Amado}, {Montes}, {Jeffers}, {Azzaro}, {B{\'e}jar}, {Chaturvedi}, {Henning}, {K{\"u}rster}, \& {Pall{\'e}}}]{Reiners22}
{Reiners}, A., {Shulyak}, D., {K{\"a}pyl{\"a}}, P.~J., {et~al.} 2022, \aap, 662, A41

\bibitem[{{R{\'e}ville} {et~al.}(2015){R{\'e}ville}, {Brun}, {Matt}, {Strugarek}, \& {Pinto}}]{Reville2015}
{R{\'e}ville}, V., {Brun}, A.~S., {Matt}, S.~P., {Strugarek}, A., \& {Pinto}, R.~F. 2015, \apj, 798, 116

\bibitem[{{Robinson} {et~al.}(2021){Robinson}, {Espaillat}, \& {Owen}}]{Robinson2021}
{Robinson}, C.~E., {Espaillat}, C.~C., \& {Owen}, J.~E. 2021, \apj, 908, 16

\bibitem[{Serna {et~al.}(2021)Serna, Hernandez, Kounkel, Manzo-Mart{\'{\i}}nez, Roman-Lopes, Rom{\'{a}}n-Z{\'{u}}{\~{n}}iga, Batista, Pinz{\'{o}}n, Calvet, Brice{\~{n}}o, Tapia, Su{\'{a}}rez, Ram{\'{\i}}rez, Stassun, Covey, Vargas-Gonz{\'{a}}lez, \& Fern{\'{a}}ndez-Trincado}]{Serna2021}
Serna, J., Hernandez, J., Kounkel, M., {et~al.} 2021, The Astrophysical Journal, 923, 177

\bibitem[{Serna {et~al.}(2024)Serna, Pinzón, Hernández, Manzo-Martínez, Mauco, Román-Zúñiga, Calvet, Briceño, López-Valdivia, Kounkel, Stringfellow, Stassun, Pinsonneault, Adame, Cao, Covey, Bayo, Roman-Lopes, Nitschelm, \& Lane}]{Serna2024}
Serna, J., Pinzón, G., Hernández, J., {et~al.} 2024, The Astrophysical Journal, 968, 68

\bibitem[{Smith {et~al.}(2023)Smith, Gillen, Hodgkin, Alves, Anderson, Battley, Burleigh, Casewell, Gill, Goad, Henderson, Jenkins, Kendall, Moyano, Ramsay, Tilbrook, Vines, West, \& Wheatley}]{Smith2023}
Smith, G.~D., Gillen, E., Hodgkin, S.~T., {et~al.} 2023, Monthly Notices of the Royal Astronomical Society, 523, 169

\bibitem[{{Steiner} {et~al.}(2021){Steiner}, {Gehrig, L.}, {Ratschiner, B.}, {Ragossnig, F.}, {Vorobyov, E. I.}, {G\"udel, M.}, \& {Dorfi, E. A.}}]{Steiner21}
{Steiner}, D., {Gehrig, L.}, {Ratschiner, B.}, {et~al.} 2021, A\&A, 655, A110

\bibitem[{{Venuti} {et~al.}(2017){Venuti}, {Bouvier}, {Cody}, {Stauffer}, {Micela}, {Rebull}, {Alencar}, {Sousa}, {Hillenbrand}, \& {Flaccomio}}]{Venuti2017}
{Venuti}, L., {Bouvier}, J., {Cody}, A.~M., {et~al.} 2017, \aap, 599, A23

\bibitem[{{Watson} {et~al.}(2016){Watson}, {Calvet}, {Fischer}, {Forrest}, {Manoj}, {Megeath}, {Melnick}, {Najita}, {Neufeld}, {Sheehan}, {Stutz}, \& {Tobin}}]{Watson2016}
{Watson}, D.~M., {Calvet}, N.~P., {Fischer}, W.~J., {et~al.} 2016, \apj, 828, 52

\bibitem[{{Weber} \& {Davis}(1967)}]{Weber67}
{Weber}, E.~J. \& {Davis}, Leverett, J. 1967, \apj, 148, 217

\bibitem[{{Zanni} \& {Ferreira}(2011)}]{Zanni2011}
{Zanni}, C. \& {Ferreira}, J. 2011, \apjl, 727, L22

\bibitem[{{Zanni} \& {Ferreira}(2013)}]{Zanni13}
{Zanni}, C. \& {Ferreira}, J. 2013, \aap, 550, A99

\end{thebibliography}
%

\appendix

\section{Comparison between different APSW models}\label{sec:comp_wind}

\begin{figure}
    \centering
         \resizebox{\hsize}{!}{\includegraphics{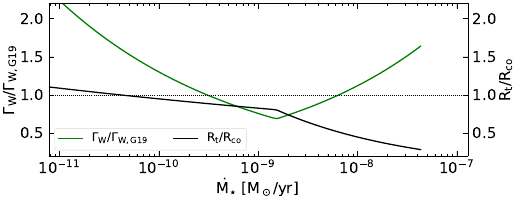}}
    \caption{
    Comparison of the APSW torque formulation used in this study, $\Gamma_\mathrm{W}$ (\equo{eq:wind}), with that in \cite{Gallet19}, $\Gamma_\mathrm{W,G19}$, for a range of accretion rates.  In addition, the value of $R_\mathrm{t}/R_\mathrm{co}$ is shown.  The stellar parameters are: $M_\star = 0.7~\mathrm{M_\odot}$, $B_\star = 0.5$~kG, $P_\star = 2$~days, and $\tau_\mathrm{age}=2$~Myr.
    The horizontal line separates the regions in which the respective wind models produce stronger torques.  
    }
    \label{fig:WG19}
\end{figure}

To assess the impact of specific APSW prescriptions on our results we compare the relative difference between our adopted APSW torque formulation \citep[][]{Ireland2022} and that adopted by \cite{Gallet19} (see \fig{fig:WG19}).
At both high and low accretion rates, $\Gamma_\mathrm{W}$  could be larger by a factor of roughly two compared to previous models (for the used parameters in \fig{fig:WG19}).  
The presence of an accretion disk affects the dependence of $r_\mathrm{A}$ on $R_\mathrm{t}$ and $\Mdot$ \citep[e.g.,][]{Ireland21} and explains the break in $\Gamma_\mathrm{W}/\Gamma_\mathrm{W,G19}$.
For high accretion rates or low field strengths (with $R_\mathrm{t}$ calculated using the first term of \equo{eq:Rt}), the Alfvén radius scales as $r_\mathrm{A}\propto \Upsilon_\star^{m_\mathrm{A}} R_\mathrm{t}^{2 m_\mathrm{\phi,1} m_\mathrm{A}}\propto \Mdot^{-0.088}$, a much weaker accretion rate dependence than other studies \citep[e.g.,][with $r_\mathrm{A}\propto \Mdot^{-0.218}$]{Gallet19} resulting in larger values of $r_\mathrm{A}$ (possibly exceeding $50 R_\star$) and $\Gamma_\mathrm{W}$.   
For low accretion rates (with $R_\mathrm{t}$ calculated using the second term of \equo{eq:Rt}), we find $r_\mathrm{A}\propto \Mdot^{-0.335}$ resulting in, again, larger values of $r_\mathrm{A}$ and $\Gamma_\mathrm{W}$ compared to \cite{Gallet19}.
However, at intermediate accretion rates close to the break in the truncation radius $\Gamma_\mathrm{W}/\Gamma_\mathrm{W,G19}<1$. 
Within this formulation of the truncation radius, large values of up to $50 R_\star$ are possible.
Whether such large values of the Alfvén radius are plausible is an open question.  
While simulations of \cite{Pantolmos20} and \cite{Ireland2022} occasionally show these values, the parameter ranges used in those studies do not completely overlap with those used in this work, especially high accretion rates at low magnetic field strengths (see \app{sec:paraIreland}).  
In addition, previous studies show significantly smaller values of $r_\mathrm{A}\lesssim20R_\star$ \citep[e.g.,][]{Matt2012APSW}.

\begin{figure}
    \centering
         \resizebox{\hsize}{!}{\includegraphics{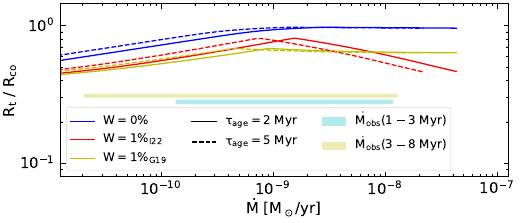}}
    \caption{
    Same as \fig{fig:theo_age}.
    In addition, we use the expression for the APSW torque presented in \cite{Gallet19} (yellow lines).
    }
    \label{fig:theoG19}
\end{figure}

To study the sensitivity of our results on the APSW model, we have recalculated the equilibrium values of $\mathcal{R}$ using the APSW model from \cite{Gallet19} (\fig{fig:theoG19}).
For high accretion rates, the values of $\mathcal{R}$ are nearly independent of $\Mdot$ and the difference between the wind models is most pronounced.
For the ranges of typical T~Tauri accretion rates, however, the models produce similar average results (see also \fig{fig:WG19}).
At low accretion rates, the equilibrium values are dominated by the effect of MEs, and the strong difference in the wind torques does not significantly affect the values of $\mathcal{R}$.

\section{Parameter space in the work of Ireland et al.}\label{sec:paraIreland}

For comparison, we compare parameter ranges from the simulations of \cite{Ireland2022} with our parameter space.
In \cite{Ireland2022}, the accretion rate ranges from $10^{-11}-7\times10^{-8}~\mathrm{M_\odot/yr}$.
The stellar magnetic field strength varies from $0.5-2.0$~kG and the values of $\mathcal{R}$ range from 0.11 to 1.16.
The value of $W$, computed from their parameters as $W=\Dot{M}_\mathrm{wind}/\Dot{M}_\mathrm{acc}$ range from $0.3\%$ up to over $100\%$.
The extremely large values of $W$ are obtained for simulation in the propeller regime when the accretion rate sharply drops.
The parameter spaces from \cite{Ireland2022} match our parameters for the most part, except for the highest accretion rates of $\Mdot \approx 10^{-7}~\mathrm{M_\odot/yr}$ and the smallest magnetic field strengths of $B_\star < 0.5$~kG.
Additional simulations would be necessary to explore the full parameter space and possibly find corrections to the given relations.



\label{LastPage}
\end{document}